\documentclass[twocolumn,showpacs,preprintnumbers,amsmath,amssymb]{revtex4}

\usepackage{graphicx}
\usepackage{dcolumn}
\usepackage{bm}

\begin{document}

\title{Continuous Time Random Walks with Internal Dynamics and Subdiffusive 
Reaction-Diffusion Equations}




\author{S.~Eule, R.~Friedrich, F.~Jenko$^1$, and I.~M.~Sokolov$^2$  }
\affiliation{Institute for Theoretical Physics, University of M\"unster,
Wilhelm-Klemm-Str.~9, D-48149 M\"unster, Germany\\
$^1$Max-Planck-Institut f\"ur Plasmaphysik,
Boltzmannstr.~2, D-85748 Garching, Germany\\
$^2$Humboldt University, Newtonstr.~15, D-12489 Berlin, Germany\\
}

\date{\today}

\begin{abstract}
We formulate the generalized master equation for a class of continuous time
random walks in the presence of a prescribed deterministic evolution between
successive transitions. This formulation is exemplified by means of an
advection-diffusion and a jump-diffusion scheme. Based on this master equation,
we also derive reaction-diffusion equations for subdiffusive chemical species,
using a mean field approximation.   

\end{abstract}

\maketitle

{\em Introduction.}
While transport in equilibrium systems occurs due to 
Brownian motion of the microscopic particles following Fick's
law, systems far from equilibrium often show anomalous -- i.e., non-Fickian
and non-Gaussian -- diffusion \cite{Metzler,Shlesinger,Zaslav}. Here,
the mean square displacement of a particle is given by a power law
$\langle x^2(t)\rangle \propto t^{\alpha}$ with $\alpha\ne 1$. As a model
of such anomalous behavior the
Continuous Time Random Walk (CTRW) introduced by Montroll and Weiss in
1965 \cite{MoWe} and extended by Scher and Montroll to explain anomalous 
diffusion \cite{Scher} has
attracted much attention during the last few decades. 
The Scher-Montroll CTRW process 
corresponds to the situation in which a particle (walker) is trapped on a 
site for a time $\tau$ distributed with a power-law probability density 
$W(\tau)$, and then makes a jump to another site. In between the jumps
it neither moves, nor changes its identity, i.e.
does not exhibit any dynamics of its own. The probability distribution of 
walker's positions is determined by a generalized master equation
\cite{MontShle}, which, under certain assumptions, takes the form of a
diffusion equation with a time memory \cite{Metzler}. Even this simple model
exhibits extremely interesting properties connected with its
intrinsic non-stationarity \cite{Nonerg} and is still
under extensive investigation both theoretically and experimentally.

In many situations, however, the system undergoes additional, internal
dynamics also during waiting periods, due e.g. to a deterministic drift 
generated by an external force or due to chemical reactions among the 
diffusing particles. 
The purpose of the present Letter is to perform a treatment of such 
processes based on the derivation of a generalized master equation.
We shall argue that, generally, these types of stochastic processes can not
be captured by a subordination procedure along the lines of Fogedby 
\cite{Fogedby}. Our approach
will allow us to formulate subdiffusive reaction diffusion equations, 
which are of utmost importance for pattern formation in biological systems,
due to the ubiguity of subdiffusive transport in biological systems
(see e.g. Ref.\cite{Nec} and references therein). 
  
Our treatment extends earlier work of Shlesinger et al. 
\cite{LeWa} on L\'evy Walks, which have been developped for the
description of particle transport in chaotic and turbulent flows, 
the work of Metzler et al. \cite{MetzKlaSok}
on passively transported particles in flows
undergoing jumps with respect to the moving fluid and the work of
Eliazar and Klafter \cite{shot} on the behaviour of overdamped particles 
exposed to random impacts, a special case of a shot noise situation.

{\em Fogedby's subordination procedure.}
Generalized master (or Fokker-Planck) equations can be obtained in a
straightforward manner by the method of subordination proposed by
Fogedby \cite{Fogedby}. Here, one considers a Markov process $x(s)$
for a variable $x$ depending on an internal (operational) time
variable $s$. The corresponding probability distribution $f(x,s)$
obeys the Fokker-Planck equation
\begin{equation} \frac{\partial }{\partial s} f_0(x,s)= L_{\rm FP}
f_0(x,s)
\end{equation} where the Fokker-Planck operator $L_{FP}$ is defined,
e.g., in Ref.~\cite{Risken}.  A related stochastic process $y(t)=x(s(t))$
is created by a time transformation $s(t)$ which is assumed to be an
independent random process with non-negative increments. It can be
shown that the probability distribution of the process $y(t)$,
$f(y,t)$, obeys a generalized Fokker-Planck equation,
\begin{equation}\label{sub} \frac{\partial }{\partial t} f(y,t)=
\int_0^t dt'\,Q(t-t')\,L_{\rm FP} f(y,t')\,,
\end{equation} whose solution reads
$f(y,t)=\int_0^t\,p(s,t)\,f_0(y,s)\,ds$ \cite{Fogedby}.
In particular, it was shown that
the quantity $p(s,t)$ is the probability distribution of the process
$s(t)$ which
is related to the kernel $Q(t)$ by the relation
\begin{equation}\label{subQ}
  \frac{\partial}{\partial t} p(s,t)=-\int_0^t dt'\,Q(t-t')\,
  \frac{\partial}{\partial s} p(s,t')\,.
\end{equation} 
The choice $Q(t-t')\propto (t-t')^{\delta-1}$ leads
to fractional equations \cite{Metzler}. Independently of the work by
Fogedby, one of us (I.S.) arrived at a similar conclusion for a much
broader class of stochastic processes and clarified the meaning of the
kernel $Q(t-t')$ \cite{Sokolov1}.
 
{\em Examples of generalized master equations.}
Due to various applications briefly mentioned above, the
case of anomalous diffusion in the presence of an additional 
deterministic process
is of significant interest. However, Fogedby's approach can 
only be extended to this
situation in a straightforward manner if the deterministic dynamics 
depends on the
{\em internal} time $s$, not on the {\em physical} time $t$. In the latter case 
one expects a quite different behaviour, and the
derivation of a respective generalized Fokker-Planck equation -- which is the main goal
of the present Letter -- turns out to be somewhat involved. Before we address this issue
in a general way, it is useful to first look at two concrete examples.

In the first one, we consider a particle with coordinate $q$ which is advected with 
a constant velocity $v$ and subjected to transitions occuring at randomly distributed
time intervals. The resulting generalized Fokker-Planck equation reads 
\begin{equation}
\left[\frac{\partial }{\partial t}+v\frac{\partial }{\partial q}\right]f(q,t)
=\int dt'\, Q(t-t') \,\frac{\partial^2 }{\partial q^2} f(q-v(t-t'),t')\,.
\end{equation}
In this context, we would like to point out the occurrence of a retardation effect in
the diffusion term (cp.~Refs.\cite{PRL,PRE}) which is due to the advective force acting
in {\em real} time on the anomalously diffusing particle. The above equation can be
solved by the ansatz $f(q,t)=F(q-vt,t)$. The probability distribution is
then governed by
\begin{equation}
\frac{\partial }{\partial t} F(\xi,t)=\int dt'\, Q(t-t') \frac{\partial^2 }{\partial
  \xi^2}F(\xi,t')\, 
\end{equation}
with $\xi=q-vt$. Thus, in the present case, a transition to a co-moving reference 
frame leads to a conventional CTRW for the variable $\xi$, 
in contrast to the behaviour for a
subordinated process described by Eq.~(\ref{sub}). Similar arguments have been given
in Ref.~\cite{MetzKlaSok}. However, the generalized Fokker-Planck equation proposed
therein is only valid to order $v^2$.

Let us now proceed to our second example. Here, we consider the motion of an overdamped
particle ($\dot q=-\gamma q$), subjected to random transitions with a suitably
defined waiting time distribution $W(t)$. 
The evolution equation for the probability distribution $f(q,t)$ 
describing such a process takes the form 
\begin{eqnarray}\label{dampedex}
& & \frac{\partial }{\partial t} f(q,t) =
\gamma \frac{\partial }{\partial q}qf(q,t)
\\
&+& \int_0^t dt' Q(t-t') \frac{\partial^2}{\partial q^2} 
\frac{1}{e^{-\gamma(t-t')}}
f\left( \frac{q}{e^{-\gamma(t-t')}},t' \right)\,.
\nonumber 
\end{eqnarray}
Using this equation, we can derive an expression for the moments
of this process to arbitrary order -- in the limit of long times.
Considering symmetric initial conditions one obtains
for the moments of even order the rather strange behaviour
$\langle q^{2k}(t)\rangle \approx  Q(t)$.
This behaviour differs from the one obtained for the subordinated case, in
which the moments tend to constants. 

{\em Derivation of a generalized master equation}.
Having considered two specific examples, we now describe the derivation of a
generalized master equation for the class of processes under consideration.
We assume that the variable $q$ undergoes a purely deterministic (or a Markovian
random) process which is characterized by the transition probability (or propagator)
$p(q,q';t-t')$. For the two examples considered above, we have, respectively,
$p(q,q',t-t')=\delta\left(q-q'-v(t-t')\right)$ and $p(q,q',t-t')=\delta
\left(q-q'e^{-\gamma(t-t')}\right)$. However, the considered class also includes
diffusion processes described by a Fokker-Planck operator. We further assume that
the variable $q$ undergoes sudden transitions from state $q$ to state $q'$ after
time intervals $\tau$ which are characterized by a waiting time distribution $W(\tau)$.
Let us denote the transition probability from state $q'$ to state $q$ by $F(q,q')$.
Then the probability that such a jump between $q$ and $q'$ occurs in the time interval
$\tau$ is just $F(q,q')\,W(\tau)$. Factorization of this quantity demonstrates that
waiting times and transitions are statistically independent which is characteristic
for decoupled CTRWs.

As usual in the theory of CTRWs, the next step is to introduce the probability density
for having arrived at times $t'$ at an infinitesimal interval close to the position $q'$
shortly after a jump. The respective quantity shall be denoted by $\eta(q',t')$. The
probability density of arriving at $q$ at time $t$ after another jump is then given by
\begin{eqnarray}
\eta(q,t) &=& \int_0^t dt'\int dq'' \int dq'\,F(q,q'')\,W(t-t')
\nonumber \\ &\times& p(q'',q',t-t')\,\eta(q',t') + \delta(t)\,f(q,0) \,.
\end{eqnarray}
Here, it has been taken into account that the system evolves after the first jump according
to the process described by the propagator $p(q'',q',t-t')$, performing another jump from
$q''$ to $q$ at time $t$. The probability density $f(q,t)$ related to the probability for 
finding a particle in the interval $dq$ close to $q$ is then given by
\begin{equation}
f(q,t)=\int_0^t dt' \int dq' \,p(q,q',t-t')\,w(t-t')\,\eta(q',t')\,.
\end{equation}
After having arrived after the jump at time $t'$ in the interval close to $q'$, the position
of the particle changes according to the propagator $p$ from $q'$ to $q$. The quantity
$w(t-t')$ is the probability that no jump occurs in the time interval $t-t'$. It is related
to the quantity $W(t-t')$ according to
\begin{equation}
  w(t-t')=1-\int_0^{t-t'} d\tau\,W(\tau)\,.
\end{equation}

The operator representation $p(q,q',t-t')=\lbrace e^{L(t-t')}\rbrace_{q,q'}$ of the propagator
$p(q,q',t-t')$ allows for the determination of the Laplace transformed equations
\begin{equation}\label{gl1}
f(q,\lambda)=w(\lambda-L) \eta(q,\lambda)
\end{equation}
and
\begin{equation}\label{gl2}
\eta(q,\lambda)=f(q,0) + \int dq' F(q,q') W(\lambda-L) \eta(q',\lambda)\,.
\end{equation}
Here, the Laplace transforms of $f(q,t)$, $w(t)$, and $W(t)$ are denoted, respectively, by
$f(q,\lambda)$, $w(\lambda)$, and $W(\lambda)$. Combining both relationships, we arrive at
the generalized master equation for the process under consideration:
\begin{eqnarray}\label{centralmaster}
\lefteqn{\left[\frac{\partial }{\partial t}-L\right]f(q,t) =
\int_0^t dt' \,Q(t-t') } 
\\
&&
\times\int dq'\, [F(q,q')-\delta(q-q')]~e^{L(t-t')}\,f(q',t')\,.
\nonumber
\end{eqnarray}
Here, we have introduced the kernel $Q(t-t')$ which is defined by its
Laplace transform \cite{MontShle}
\begin{equation}
Q(\lambda)=\frac{\lambda\,W(\lambda)}{1-W(\lambda)}\,.
\end{equation}
For example, an exponential waiting time distribution $W(t)=\Gamma e^{-\Gamma t}$ leads
to the kernel $Q(t)=\delta(t)$. The representation of $Q(t)$ in terms of the waiting time
is necessary in order to obtain generalized master equations
which define a nonnegative probability density $f(q,t)$ for all times $t$ \cite{Sokolov1}. 

At this point, it is useful to recall that, allowing only for nearby jumps, the transition
probability $F(q,q')$ in the master equation yields a generalized Fokker-Planck equation
\cite{BarkMetz}. In our case, it is  of the form:
\begin{equation}\label{central}
\left[\frac{\partial }{\partial t}-L\right]f(q,t)=
\int_0^t dt' {\it L_{\rm FP}} Q(t-t')e^{L(t-t')} f(q,t')\, .
\end{equation}
Here, the two operators $L$ and $L_{\rm FP}$ arise. The Fokker-Planck operator $L_{\rm FP}$
is connected with the transition probability $F(q,q')$ of the time-random sudden jumps
\cite{BarkMetz}, whereas the operator $L$ is connected with the propagator $p(q,q',t,t')$
and describes the continuously evolving process. This is the desired Fokker-Planck-type
equation for the class of stochastic processes under consideration -- a key result of the
present paper, representing a nontrivial generalization of Eq.~(\ref{sub}) with the operator
$L$ --  describing the time evolution of the system between successive jumps -- entering on
both sides of the equation.

{\em Analytic solutions.}
At first glance, Eq.~(\ref{central}) appears to be rather complicated, maybe not analytically
tractable. However, as the above examples show, analytical solutions are indeed possible.
In this context, we would like to briefly mention the following strategy 
to solve this
generalized Fokker-Planck equation. It turns out to be convenient 
to switch to a kind of
interaction picture by employing the ansatz
\begin{equation}
f(q,t)=e^{Lt}g(q,t)\, .
\end{equation}
Then, the resulting problem to be solved is 
\begin{equation}
\frac{\partial }{\partial t}g(q,t)=
 e^{-Lt}{\it L_{FP}}e^{Lt} \int_0^t dt'\,Q(t-t')\,g(q,t')\,.
\end{equation}
Provided the two Fokker-Planck operators $L$ and $L_{FP}$ commute, 
we (only) have to solve
the simpler problem
\begin{equation}
\frac{\partial }{\partial t} g(q,t)= L_{FP} \int_0^t dt'\, Q(t-t')\,g(q,t')\,.
\end{equation}
Comparing this result with the Fokker-Planck equation (\ref{sub}), it is evident that
$g(q,t)$ describes a subordinated process.

The first of the above mentioned examples falls into this catagory. 
As a further example, we would like
to consider the two operators $L=Q_1\frac{\partial^2}{\partial q^2}$ and
$L_{\rm FP}=Q_0\frac{\partial^2}{\partial q^2}$ describing pure diffusion interrupted by
randomly distributed sudden transitions, i.e., a jump-diffusion process. The corresponding
probability distribution takes the form
\begin{equation}
f(q,t)= \int_0^\infty ds\, p(s,t) \,\exp\left[-\frac{q^2}{2(Q_0s+Q_1t)}\right]\,.
\end{equation}
Calculating the even moments we find that their 
long term behaviour is dominated by the
scaling behaviour $\langle q(t)^{2k}\rangle \propto t^k$. The additional
jumps result in subdominated scalings.

{\em Subdiffusive reaction-diffusion equations}.
The master equation just derived can also be used to obtain reaction-subdiffusion
equations -- which happens to be a particularly interesting application. 
The subdiffusive transport is modeled
by the introduction of a random waiting times between the jumps of particles. 
The reaction among the chemical
species evolves however continuously in time, changing the concentrations also
between the jumps. Our derivation starts from extending the
procedure leading to Eq.~(\ref{centralmaster}) to multiple dimensions, leading
to a master equation for a multi-dimensional state vector, and 
the reaction-subdiffusion equations are then derived 
along the lines of the derivation of the reaction-diffusion equations from 
the usual Fokker-Planck equations.

We partition real space into small compartments labeled by the index $i$ and
consider $N$ different chemical species (labelled by $\alpha=1,..,N$) which locally 
react according to the deterministic reaction kinetics
\begin{equation}\label{kin}
\dot c_{\alpha,i} = R_\alpha(c_{\alpha,i})\, .
\end{equation}
Further, we allow for diffusive transitions between neighbouring
cells. For the sake of simplicity, we assume that the transitions of
all particles occur simultaneously (``global
update'') at random times, characterized by the waiting time distribution $W(t)$. The
situations with independent particle jumps (``local update'') are also
treatable but lead to considerably more complicated
calculations. Lumping all concentrations $c_{\alpha,i}$ into the state
vector ${\bf c}$, we can formulate the generalized master equation:
\begin{eqnarray}
\frac{\partial }{\partial t} f({\bf c},t) &=&
-\frac{\partial }{\partial {\bf c}} {\bf R}({\bf c} ) 
f({\bf c},t)
+ \int_0^t dt' Q(t-t') 
\\
&\times&
\int d{\bf c}' 
\int d{\bf c}''
\left[ F({\bf c},{\bf c}')-\delta({\bf c}-{\bf c}') \right]
\nonumber \\
&\times&
\delta({\bf c}'-{\bf G}({\bf c}'',t-t'))
f({\bf c}'',t')\, .
\nonumber 
\end{eqnarray}
Here the function ${\bf G}({\bf c}',t-t')$ is the solution of the kinetic equation
(\ref{kin}) with the initial condition ${\bf G}({\bf c}',0)={\bf c}'$.

The desired subdiffusive 
reaction-diffusion equation is an evolution equation for the mean
value of the quantity ${\bf c}$,
${\bf C}=\int d{\bf c}\, {\bf c}\, f({\bf c},t)$. It reads
\begin{eqnarray}
& & \frac{\partial }{\partial t}{\bf C} = \int {\bf R}({\bf c}) f({\bf c},t)
+\int_0^t dt'\,Q(t-t')
\nonumber \\
& & \times\left[\right.
\int d{\bf c}\, d{\bf c}'\,{\bf c}\, F({\bf c},{\bf G}({\bf c}',t-t'))\,f({\bf c}',t')
\nonumber \\
& & -\int d{\bf c}'\,{\bf G}({\bf c}',t-t'))\,f({\bf c}',t') \left.\right]\,.
\end{eqnarray}
The mean field approximation $ \int d{\bf c}\, H({\bf c})\,f({\bf c},t) \approx H({\bf C})$
leads then to a closed equation for the mean concentrations. Allowing 
only for nearest-neighbour transitions, we obtain the reaction-subdiffusion equation 
\begin{eqnarray}\label{genmar}
\frac{\partial }{\partial t}{\bf C}({\bf x},t) &=& {\bf R}({\bf C}({\bf x},t))
\\
&+& D \Delta_x \int_0^t dt' Q(t-t') {\bf G}\left({\bf C}({\bf x},t'),t-t'\right)\,.
\nonumber 
\end{eqnarray}

There have been several attempts to establish reaction-diffusion equations for
subdiffusive chemical species (see \cite{Saques} and references therein). 
Ad hoc formulations can be obtained 
from Eq.~(\ref{genmar}) by replacing $G(C({\bf x},t'),t-t')$ with $C({\bf x},t')$.
However, such equations show serious inconsistencies; in some cases, even 
mass conservation is
violated \cite{Saques}. The correct way to proceed was discussed in \cite{Saques}
for a special example of a linear bimolecular reaction scheme 
$\mathrm{A} \rightleftharpoons \mathrm{B}$.  
This case is contained in our general subdiffusive
reaction-diffusion equation as follows. If we take the
reaction rates ${\bf R}=M{\bf C}$ to be linear in ${\bf C}$, we obtain
${\bf G}\left({\bf C}({\bf x},t'),t-t'\right)=e^{M(t-t')} {\bf C}({\bf
  x},t')$. For a two-component vector ${\bf C}({\bf x},t)$ 
we end up directly with the system considered in \cite{Saques}.
Since the particles in this scheme do not interact, the global update
assumption leads to the same results as the independent particle motion considered in
Ref.\cite{Saques}. An important application is the treatment of radiaoctive 
decay in flows through porous media \cite{Zoia}.

Finally, as a nonlinear example, we consider the subdiffusive 
Fisher-Kolmogorov-Petrovskii-Piskunov (FKPP) equation, which has been
proposed as a model equation for the propagation of favorable gene in a population \cite{Fisher}. 
The diffusive version of the FKPP equation is one of the basic equations for the investigation of 
reaction fronts in nonlinear reaction kinetics. Our subdiffusive version for the
$\mathrm{A}+\mathrm{A} \rightleftharpoons \mathrm{A}$ reaction reads
\begin{eqnarray}
\lefteqn{
\frac{\partial }{\partial t}C(x,t)=C(x,t)\,(1-C(x,t))
}\\
&&+
\frac{\partial^2}{\partial x^2} \int_0^t dt'\, Q(t-t') \,\frac{C(x,t')}
{[1-e^{-(t-t')}]C(x,t')+e^{-(t-t')}}\,.
\nonumber 
\end{eqnarray}
Note the emergence of a nonlinear
diffusion term in addition to a temporal memory. 
We stress that the equation for a reversible reaction under 
global update differs from the one obtained in 
Ref.\cite{Froem} for the locally updated irreversible $\mathrm{A}+\mathrm{B} \rightarrow
2\mathrm{B}$ reaction scheme.   

{\em Summary.} We have considered stochastic processes which are partly
generated by Markovian processes and which, {\em additionally}, 
are subjected to the
impact of fluctuations randomly occuring in time. These impacts are
treated in the framework of continuous time random walk processes by a
waiting time distribution. We have derived the generalized master 
equation for this class of processes and have been able to formulate 
subdiffusive reaction-diffusion systems also for nonlinear reaction
kinetics, which are of relevance for pattern formation in biological systems.

\end{document}